\documentclass[lettersize,journal]{IEEEtran}
\usepackage{amsmath,amsfonts}
\usepackage{algorithmic}
\usepackage{algorithm}
\usepackage{array}
\usepackage{textcomp}
\usepackage{stfloats}
\usepackage{url}
\usepackage{verbatim}
\usepackage{graphicx}
\usepackage{cite}

\usepackage{amsfonts} 
\usepackage{physics} % 长竖线 
\usepackage{algorithmic}  % 算法框架
\usepackage{algorithm} 
\usepackage{graphicx}
\usepackage{color} % 字体颜色
\usepackage{bm}
\usepackage{epstopdf} 
\usepackage{subfigure}
\usepackage{booktabs}  % table
\usepackage[dvipsnames]{xcolor} % 更多字体颜色
\usepackage{cite} % 多项引用
\usepackage{balance}
\usepackage{arydshln} %  绘制分块矩阵
\usepackage{mathrsfs} % 花体
\usepackage{cuted} % 单栏公式

\usepackage{multirow} % 表格中合并多行
\usepackage{makecell} % 表格内换行

\begin{document}
	
	\title{Weighted MMSE Precoding for Constructive Interference Region}
	
	\author{Yafei Wang, \textit{Student Member}, \textit{IEEE}, Wenjin Wang, \textit{Member}, \textit{IEEE}, Li You, \textit{Senior Member}, \textit{IEEE}, \\ Christos G. Tsinos, \textit{Senior Member}, \textit{IEEE}, Shi Jin, \textit{Senior Member}, \textit{IEEE}}
	
	% \markboth{IEEE WIRELESS COMMUNICATIONS LETTERS}%
	
	% 页尾
	
	\maketitle

	\begin{abstract}
		In this paper, we propose a symbol-level precoding (SLP) design that aims to minimize the weighted mean square error between the received signal and the constellation point located in the constructive interference region (CIR). Unlike most existing SLP schemes that rely on channel state information (CSI) only, the proposed scheme exploits both CSI and the distribution information of the noise to achieve improved performance.
		We firstly propose a simple generic description of CIR that facilitates the subsequent SLP design.
 		Such an objective can further be formulated as a nonnegative least squares (NNLS) problem, which can be solved efficiently by the active-set algorithm. Furthermore, the weighted minimum mean square error (WMMSE) precoding and the existing SLP can be easily verified as special cases of the proposed scheme. Finally, simulation results show that the proposed precoding outperforms the state-of-the-art SLP schemes in full signal-to-noise ratio ranges in both uncoded and coded systems without additional complexity over conventional SLP.
	\end{abstract}
	

	% For peer review papers, you can put additional information on the cover
	% \fi
	\IEEEpeerreviewmaketitle

	\section{Introduction}
	\IEEEPARstart{I}{n} 
	multiuser multi-input multi-output (MU-MIMO) downlink transmission, precoding is employed to suppress user interference and to obtain spatial multiplexing and array gain. 
	Linear precoding, e.g., zero-force (ZF) and weighted minimum mean square error (WMMSE) \cite{1040325, 974266}, constructs the precoding matrix based on channel state information. Although linear precoding is widely adopted due to its low complexity, it cannot approach theoretical system capacity.
	In contrast, nonlinear precoding further improves the performance by designing the precoding scheme with input data, e.g., dirty paper coding (DPC) \cite{1056659}, vector
	perturbation (VP) precoding \cite{1413598},  and symbol-level-precoding (SLP) \cite{ 9035662, Li2021,7103338,9593254,7417066,8466792,8299553,8477154}. 
	
	Among wireless nonlinear precoding, SLP exploits interference with the transmitted data symbols and their corresponding constellations to design precoding schemes symbol-by-symbol \cite{ 9035662, Li2021,7103338,9593254,7417066,8466792,8299553,8477154}. 
	The concept of constructive interference (CI) and destructive (DI) was proposed in \cite{4801492}, based on which the CI region (CIR) was defined, and precoding was therefore specifically optimized for phase-shift keying (PSK) in \cite{7042789,7103338}. 
	The definition of CIR in PSK was further extended to quadrature amplitude modulation (QAM)  \cite{8299553,8477154,7417066}, where the inner constellations were fixed, and outer constellations could be expanded. Furthermore, SLP is proved to be the generalization of ZF precoding in \cite{8466792} and \cite{Li2021}, and it is also represented by the form of symbol-perturbed ZF precoding in \cite{8462190} and \cite{article}.
	
	Although the existing SLP solutions achieve significant performance gains over ZF precoding in the high SNR regime by exploiting CI, they are inferior to WMMSE precoding in the low signal-to-noise ratio (SNR) regime {as they only control interference but ignore the impact of the noise, while WMMSE utilizes distribution information of the noise} {\cite{9593254, 4712693, 1391204}}.
	This raises a key question: \textit{How to exploit both CI and the {distribution information of the noise} to optimize precoding design in full SNR ranges?}
	{In this paper, we try to provide an answer. Firstly, we propose a simple generic description of CIR, which can facilitate the related SLP designs. Then, {we develop CI-WMMSE precoding for the first time}, which minimizes the weighted mean square error between the received signal and the {target signal} in CIR. {CI-WMMSE is expected to perform better since it synthetically utilizes the noise distribution and CIR.} There is a remarkable conclusion that the WMMSE and the existing SLP can be verified as special cases of CI-WMMSE. Based on our CIR description, CI-WMMSE is formulated as a nonnegative least squares (NNLS) problem, which can be efficiently solved \cite{lawson1995solving}.} {Finally, simulation results show the superiority of CI-WMMSE in full SNR ranges in both coded and uncoded systems}.
	
	\section{System Model}
	
	Consider a downlink system where a $N$-antenna base station (BS) transmits information to $K$ single-antenna user equipments (UE). The channel between the BS and the $k$-th UE is denoted as ${\bf h}_k\in{\mathbb{C}}^{N\times 1}$. The channel matrix ${\bf H} = \left[{\bf h}_1,{\bf h}_2 ..., {\bf h}_K\right]^T$ is assumed to be available at the BS. At a symbol duration, $K$ independent QAM or PSK symbols are intended to be transmitted to $K$ UEs, and the symbols are denoted as ${\bf s} = \left[s_{1},s_{2} ..., s_{K}\right]^T\in{\mathbb{C}}^{K\times 1}$,
	where $s_{k}$ is the symbol to be transmitted to the $k$-th UE. The symbol vector ${\bf s}$ is mapped to the transmit vector ${\bf u}\in{\mathbb{C}}^{N\times 1}$ by applying the symbol-level precoder {represented by} ${\rm SLP}\left(\cdot\right)$, which can be expressed as
	\begin{align}
	{\bf u} = {\rm SLP}\left({\bf s}, {\bf H}, \sigma^2\right),
	\end{align}
	{where $\sigma^2$ denotes the variance of the additive noise $n_k$ following zero-mean complex Gaussian distribution at the $k$-th UE.} The received signal at the $k$-th UE is given by
	\begin{equation}
	\label{E1}
	{y}_k = {\bf h}^T_k{\bf u} + n_k,\;\forall k \in {\mathcal{ K}},
	\end{equation}
	{where ${\mathcal{ K}}=\left\{1, 2, ..., K\right\}$. When multi-level modulations {(e.g., 16QAM and 64QAM)} are employed, the received signals require to be scaled for correct demodulation, i.e.,
		\begin{align}
		\hat{y}_k = \frac{a_k}{\gamma}\left({\bf h}^T_k{\bf u} + n_k\right),
		\label{y_k}
		\end{align}
		{where $a_k\in\mathbb{C}$ is a nonzero complex that denotes the pre-determined processing coefficient at $k$-th UE, and ${\gamma}\in\mathbb{R}$ is a gain factor in the precoding scheme which scales the signal
	 	to satisfy the symbol-level transmit power constraint $\|{\bf u}\|^2_2\leq P_T$, where $P_T$ is the transmit power budget. By omitting the noise component of ${\hat y}_k$, $(a_k/\gamma){\bf h}^T_k{\bf u}$ is termed as the noise-free received signal.}
		
		\begin{figure}[t]
			\centering
			\includegraphics[width=3in]{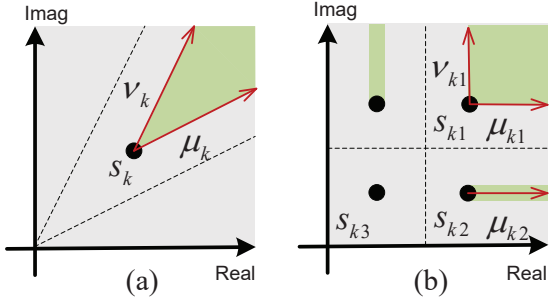}
			\caption{CIRs and their boundary vectors of (a) PSK and (b) QAM.}
			\label{CI_MMSE_CIR}
		\end{figure}
		
		\section{WMMSE Criterion for Constructive Interference Region}
		
		In conventional SLP transmission, CI is defined as the interference that pushes the noise-free received
		signal further away from the decision boundaries. 
		CI region (CIR) is the modulation-specific
		region where the interference component of the received signal is CI \cite{9035662,8299553,Li2021}. In this Section, we firstly propose a simple parameterized description of CIR, which can facilitate subsequent SLP design.

		\subsection{A Simple Generic Description of CIR}
				\label{CIR description}
		The CIR for PSK and QAM constellations, {whose geometric definitions have been well investigated \cite{Li2021},} are shown in the green areas of Fig. \ref{CI_MMSE_CIR}, where the dotted lines represent the decision boundaries based on the maximum likelihood (ML) decision rule \cite{8299553}. {The CIRs, composed of two or fewer boundaries (including those of PSK and QAM), can be described by the transmitted symbol and its CIR boundary parameters.} Specifically, the CIR belongs to $s_k$ in Fig. \ref{CI_MMSE_CIR} can be expressed as 
		\begin{align}
		\mathcal{D}_k = \left\{{{{\tilde {s}}_k}}\left|
		{{\tilde {s}}_k} = s_k + \delta_{\mu_k}{{\mu}}_{k}+\delta_{\nu_k}{{\nu}}_{k}
		\right.,\ \delta_{\mu_k},\delta_{\nu_k}>0\right\},
		\label{CIR-D}
		\end{align}
		{where ${{\mu}}_{k}$ and ${{\nu}}_{k}$ are two normal boundary parameters of the CIR belonging to $s_k$, which can be easily obtained from $s_k$.}
		We define the following diagonal matrixes
		\begin{align}
		\begin{split}
		{\bf M}_R &= {\rm diag}\{\real({{\mu}}_{1}),...,\real({{\mu}}_{K})\},\\{\bf M}_I &= {\rm diag}\{\imaginary({{\mu}}_{1}),...,\imaginary({{\mu}}_{K})\},\\
		{\bf N}_R &= {\rm diag}\{\real({{\nu}}_{1}),...,\real({{\nu}}_{K})\},\\
		{\bf N}_I &= {\rm diag}\{\imaginary({{\nu}}_{1}),...,\imaginary({{\nu}}_{K})\},
		\end{split}
		\end{align}
		and construct a block diagonal matrix ${\boldsymbol{\Lambda}} = 
		\begin{bmatrix}
		{\bf M}_R & {\bf N}_R\\
		{\bf M}_I & {\bf N}_I\\		
		\end{bmatrix}.$
		{Since real and imaginary parts of ${{\tilde {s}}_k}\in\mathcal{D}_k$ can be expressed as
			\begin{align}
			\begin{split}
			\real({{\tilde {s}}_k}) &= \real(s_k)+ \delta_{\mu_k}\real({{\mu}}_{k})+\delta_{\nu_k}\real({{\nu}}_{k}),\\
			\imaginary({{\tilde {s}}_k}) &= \imaginary(s_k)+ \delta_{\mu_ k}\imaginary({{\mu}}_{k})+\delta_{\nu_ k}\imaginary({{\nu}}_{k}),
			\end{split}
			\end{align} 
			we further define ${\boldsymbol{\delta}} = \left[\delta_{\mu_1}, ..., \delta_{\mu_K}, \delta_{\nu_1}, ..., \delta_{\nu_K} \right]^T$, based on which the real and imaginary parts of signal vector $\tilde{\bf s} = \left[{\tilde{s}}_{1},{\tilde{s}}_{2}, ..., {\tilde{s}}_{K}\right]^T$ with the constraint
			\begin{align}
			s.t.~ & {\tilde{s}}_k\in \mathcal{D}_k,\;\forall k\in{\mathcal{ K}},
			\label{CIR constraint}
			\end{align}
			can be equivalently expressed as 
			\begin{align}
			\begin{bmatrix}
			\real({\tilde {\bf s}})\\
			\imaginary({\tilde {\bf s}})
			\end{bmatrix}
			=
			\begin{bmatrix}
			\real({ {\bf s}})\\
			\imaginary({ {\bf s}})
			\end{bmatrix}
			+
			{\boldsymbol{\Lambda}}{\boldsymbol{\delta}},\ {\boldsymbol{\delta}}{\succeq}{\bf 0}.
			\label{equivalent expression}
			\end{align}
			Notice that such a simple and clear generic description can significantly facilitate the following derivations.}
		
		\begin{figure}[t]
			\centering
			\includegraphics[width=3in]{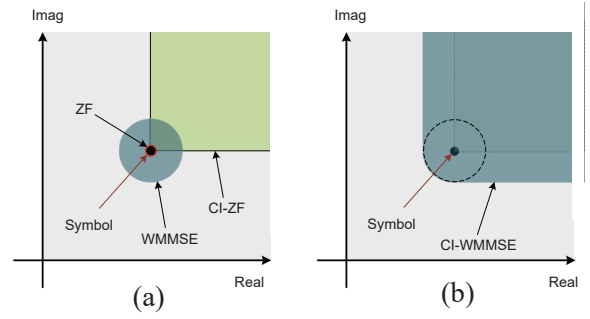}
			\caption{Distribution area of noise-free received signal for (a) conventional precoding schemes and (b) CI-WMMSE ({our SLP scheme}).}
			\label{CI_MMSE_CIR_transfom}
		\end{figure}
		
		\subsection{Constructive-Interference WMMSE Precoding}
		{As shown in the red area in Fig. \ref{CI_MMSE_CIR_transfom}(a), ZF precoding  restrains the noise-free received signal at the position of transmitted symbol $s_k$, aiming to suppress all interference. Over the low SNR regime, since ZF precoding ignores the impact of noise, it underperformed WMMSE precoding \cite{Li2021}. Given the noise distribution, WMMSE allowed the noise-free received signal to be distributed around $s_k$ as the dark area in Fig. \ref{CI_MMSE_CIR_transfom}(a) to minimize the weighted mean square error (MSE) between $\hat{y}_k$ and $s_k$.
		As the green area in Fig. \ref{CI_MMSE_CIR_transfom}(a) shows, CI-ZF precoding (i.e., the conventional SLP \cite{7103338, Li2021, 8466792, 8815429, article}) allows the noise-free received signal to distribute in the CIR, which relaxes the constraints on interference. 
		Similar to ZF, CI-ZF also ignores the impact of the noise and thus can only achieve performance gain in high SNR regimes. 
		One potential way to improve precoding performance is to exploit both the benefit of CIR and the noise distribution.} 
		To this end, we investigate the CI-WMMSE precoding, aiming to minimize the weighted MSE between $\hat{y}_k$ and the expected received signal ${\hat{s}}_k$ located in CIR $\mathcal{D}_k$.
		Define ${\bf u}=\gamma {\bf x}$, where $\gamma=\sqrt{\frac{P_T}{\left \|{\bf x}\right\|_2^2}}$ is the symbol-level power scaling factor, and ${\bf x}$ is the transmit signal without power constraint. 
		Then \eqref{y_k} can be rewritten as
		\begin{align}
		\begin{split}
		\hat{y}_k
		= a_k{\bf h}_k{\bf x} + \frac{a_k n_k}{\gamma}.
		\end{split}
		\end{align}
		Thus, the problem of CI-WMMSE can be expressed as
		\begin{align}
		\begin{split}
		\min\limits_{{\bf x}, {\tilde{\bf s}}} \ &\mathbb{E}_{{\bf n}}\left\{\left \|{\boldsymbol{\Omega}}^{\frac{1}{2}}\left[{\bf A}\left({\bf H}{\bf x}+\frac{{\bf n}}{\gamma}\right)-{\tilde{\bf s}}\right]\right\|_2^2\right\}	
		\\
		s.t.~ & {\tilde{s}}_k\in \mathcal{D}_k,\;\forall k\in{\mathcal{ K}},
		\end{split}
		\label{CI-MMSE0}
		\end{align}	
		where ${\bf A} = {\rm diag}\{a_{1},a_{2},...,a_{K}\}$, ${\boldsymbol{\Omega}}={\rm diag}\{\omega_{1},\omega_{2},...,\omega_{K}\}$, {${\tilde{\bf s}}$ denotes the vector of signals located in CIR defined in \eqref{CIR constraint} and \eqref{equivalent expression}}, {$\omega_{k}$ represents the weight belonging to the MSE of $k$-th UE}, {and $\mathbb{E}_{{\bf n}}\left\{\cdot\right\}$ denotes taking an expectation over ${\bf n}$}.
		As Fig. \ref{CI_MMSE_CIR_transfom}(b) shows, CI-WMMSE relaxes the constraint in WMMSE that the expected received signal should be around the transmit symbol. 
		By taking expectation over ${\bf n}$, \eqref{CI-MMSE0} becomes
		\begin{align}
		\begin{split}
		\min\limits_{{\bf x}, {\tilde{\bf s}}} \
		&\left \|{\boldsymbol{\Omega}}^{\frac{1}{2}}\left({\bf A}{\bf H}{\bf x}-{\tilde{\bf s}}\right)\right\|_2^2  +\frac{\sigma^2{\rm Tr}\left(\boldsymbol{\Omega}{\bf A}{\bf A}^H\right)}{P_T}\left \|{\bf x}\right\|_2^2
		\\
		s.t.~ & {\tilde{s}}_k\in \mathcal{D}_k,\;\forall k\in{\mathcal{ K}}.		
		\end{split}
		\label{CI-WMMSE0_2}
		\end{align}

		\textbf{\textit{Remark 1:}} The problems of WMMSE, MMSE, CI-ZF, and ZF precoding can be viewed as special cases of CI-WMMSE precoding.
		
		\textbf{	 \textit{1) WMMSE \& MMSE:}} 
		{If we constrain ${\tilde{\bf s}}={\bf s}$ and the precoding scheme as the linear precoding}, i.e., ${\bf x} = {\bf W}{\bf s}$, the problem \eqref{CI-MMSE0} degenertates into  
		\begin{align}
			\min\limits_{{\bf W}} \ &\mathbb{E}_{{\bf n}}\left\{\left \|{\boldsymbol{\Omega}}^{\frac{1}{2}}\left[{\bf A}\left({\bf H}{\bf W}{\bf s}+\frac{{\bf n}}{\gamma}\right)-{{\bf s}}\right]\right\|_2^2\right\}.
			\label{WMMSE}
		\end{align}
		The above problem is equivalent to the conventional WMMSE except for the definition of $\gamma$. Unlike power scaling factor $\gamma=\sqrt{\frac{P_T}{\left \|{\bf W}{\bf s}\right\|_2^2}}$ defined in \eqref{WMMSE}, conventional WMMSE defines it as $\gamma=\frac{P_T}{{\rm Tr}\left({\bf W}{\bf W}^H\right)}$ \cite{4712693}, which means the former is constrained by symbol-level power and the latter by average power. Consequently, as a special case of CI-WMMSE, problem \eqref{WMMSE} can be seen as symbol-level power constrained WMMSE. When the power allocation scheme in \cite{9593254} is employed { and the number of symbol groups involving in the scheme approaches infinity}, problem \eqref{WMMSE} with symbol-level power constraint can be easily verified to be equivalent to that with average power constraint, since the average power constraint can be regarded as a power allocation scheme between symbols.
		When  $\boldsymbol{\Omega}={\bf I}_K$, problem \eqref{WMMSE} will further degenerate into 		
		MMSE \cite{1049276}.
		
		\textbf{	 \textit{2) CI-ZF \& ZF:}} 
		When $\boldsymbol{\Omega}={\bf I}_K$ and $\sigma\to 0$,
		the left part of the cost function  in \eqref{CI-WMMSE0_2} can be seen as a constraint $\left\|{\bf A}{\bf H}{\bf x}-{\tilde{\bf s}}\right\|_2^2=0$, and the right part $\left\|{\bf x}\right\|_2^2$ becomes the primary cost function, based on which CI-WMMSE \eqref{CI-WMMSE0_2} is transformed into 
		\begin{align}
		\begin{split}
		&\min\limits_{{\bf x}} \
		\left \|{\bf x}\right\|_2^2		
		\\
		&s.t.~  {\bf h}^T_k{\bf x}\in a^{-1}_k \mathcal{D}_k,\;\forall k\in\mathcal{K},
		\end{split}
		\label{CI-ZF formulation}
		\end{align}
		which can be easily verified to be equivalent to the problem of CI-ZF in \cite{8477154}. Furthermore, {if we set} ${\bf A} = {\bf I}_K$ and ${\tilde{\bf s}}={\bf s}$, ${\bf x}$ is only subjected to the constraint $\left\|{\bf H}{\bf x}-{{\bf s}}\right\|_2^2=0$ which is the problem of ZF.

		{Considering the equivalent expression of  ${\tilde{\bf s}}$ in \eqref{equivalent expression},} the representation of \eqref{CI-WMMSE0_2} in the real numbers' domain can be written as
		\begin{align}
		\min\limits_{\bar{\bf x},{\boldsymbol{\delta}}{\succeq}{\bf 0}} \left \|{\bar{\boldsymbol{\Omega}}}^{\frac{1}{2}}\left(\bar{\bf A}\bar{\bf H}\bar{\bf x}-\left(\bar{\bf s}+{\boldsymbol{\Lambda}}{\boldsymbol{\delta}}\right)\right)\right\|_2^2  +\frac{\sigma^2{\rm Tr}\left({\bar{\boldsymbol{\Omega}}}{\bar{\bf A}}{\bar{\bf A}}^T\right)}{P_T}\left \|\bar{\bf x}\right\|_2^2,
		\label{CI-MMSE1}
		\end{align}		
		where
		\begin{align}
		\begin{split}
		&\ \ \bar{\bf s} = \begin{bmatrix}
		\real\left({\bf s}\right)\\ \imaginary\left({\bf s}\right)
		\end{bmatrix},
		\bar{\boldsymbol{\Omega}}  =
		\begin{bmatrix}
		{\boldsymbol{\Omega}}&  \\
		&{\boldsymbol{\Omega}}
		\end{bmatrix},
		\bar{\bf x} = \begin{bmatrix}
		\real\left({\bf x}\right)\\ \imaginary\left({\bf x}\right)
		\end{bmatrix},\\
		\bar{\bf A} &=
		\begin{bmatrix}
		\real\left({\bf A}\right)& -\imaginary\left({\bf A}\right)\\
		\imaginary\left({\bf A}\right)& \real\left({\bf A}\right)
		\end{bmatrix},
		\bar{\bf H} =
		\begin{bmatrix}
		\real\left({\bf H}\right)& -\imaginary\left({\bf H}\right)\\
		\imaginary\left({\bf H}\right)& \real\left({\bf H}\right)
		\end{bmatrix}.
		\end{split}
		\label{real2}
		\end{align}
		Define $\rho={\rm Tr}\left({\bar{\boldsymbol{\Omega}}}{\bar{\bf A}}{\bar{\bf A}}^T\right)$ and $\breve{\bf H} = {\bar {\bf A}}{\bar {\bf H}}$, the gradient of $\bar{\bf x}$ is given by
		\begin{align}
		\frac{\partial f}{\partial \bar{\bf x}} &=  2{\breve {\bf H}}^T{\bar{\boldsymbol{\Omega}}}\breve{\bf H}\bar{\bf x}-2\breve{\bf H}^T{\bar{\boldsymbol{\Omega}}}\left(\bar{\bf s}+{\boldsymbol{\Lambda}}{\boldsymbol{\delta}}\right)+2\frac{\sigma^2\rho}{P_T}\bar{\bf x},
		\label{gradient}
		\end{align}
		where $f$ denotes cost function in \eqref{CI-MMSE1}. Since $f$ is a convex function over the open set of variables $\bar{\bf x}$, The optimal $\bar{\bf x}^{\ast}$ can be found by setting the gradient in \eqref{gradient} to zero and can be written as
		\begin{align}
		\bar{\bf x}^{\ast} = {\breve{\bf H}}^T{\bar{\boldsymbol{\Omega}}}\left({\breve{\bf H}}{\breve{\bf H}}^T{\bar{\boldsymbol{\Omega}}}+\frac{\sigma^2 \rho}{P_T}{\bf I}_{2K}\right)^{-1}\left({\bar {\bf s}}+{\boldsymbol{\Lambda}}{\boldsymbol{\delta}}\right).
		\label{solution 1}
		\end{align}
		By replacing $\bar{\bf x}$ in \eqref{CI-MMSE1} with the formulation of $\bar{\bf x}^{\ast}$,  \eqref{CI-MMSE1} can be rewritten as \eqref{CI-WMMSE4}.
		\begin{equation}
		\min\limits_{{\boldsymbol{\delta}}{\succeq}{\bf 0}}
		\left({\bar {\bf s}}+{\boldsymbol{\Lambda}}{\boldsymbol{\delta}}\right)^T\left({\breve{\bf H}}{\breve{\bf H}}^T+\frac{\sigma^2 \rho}{P_T}{\bar{\boldsymbol{\Omega}}}^{-1}\right)^{-1}
		\left({\bar {\bf s}}+{\boldsymbol{\Lambda}}{\boldsymbol{\delta}}\right).
		\label{CI-WMMSE4}
		\end{equation}
		As a quadratic programming problem with nonnegative constraints, the above formulation is further reformulated as a much simpler NNLS problem in the next subsection.

		\textbf{\textit{Remark 2:}} The optimal solutions of WMMSE, MMSE, CI-ZF, and ZF  precoding can be viewed as special cases of CI-WMMSE precoding. 
		
		\textbf{	 \textit{1) WMMSE \& MMSE:}} 
		When ${\boldsymbol{\delta}}={\bf 0}$, the optimal solution \eqref{solution 1} degenerates into real representations of WMMSE \cite{4712693}, which further degenerates into MMSE when $\bar{\boldsymbol{\Omega}}={\bf I}_{2K}$,
		
		\textbf{	 \textit{2) CI-ZF \& ZF:}} 
		When $\boldsymbol{\Omega}={\bf I}_K$ and $\sigma\to 0$, the optimal solution \eqref{solution 1} degenerates into
		\begin{align}
		\bar{\bf x}^{\ast}_{CIZF} &= {\bar{\bf H}}^T\left({\bar{\bf H}}{\bar{\bf H}}^T\right)^{-1}\left({\bar{\bf A}}^{-1}{\bar {\bf s}}+{\bar{\bf A}}^{-1}{\boldsymbol{\Lambda}}{{\boldsymbol{\delta}}}\right),
		\label{CIZF}
		\end{align}	
		which can be easily verified as equivalent to CI-ZF in \cite{8815429}. \eqref{CIZF} will further degenerate into ZF when $\bar{\bf A} = {\bf I}_{2K}$ and ${\boldsymbol{\delta}}={\bf 0}$.

		\subsection{Low-Complexity Algorithmic Solution}
		By obtaining the upper triangular matrix ${\bf B}$ from the Cholesky decomposition
		${\bf B}^{T}{\bf B} = \left({\breve{\bf H}}{\breve{\bf H}}^T+\frac{\sigma^2 \rho}{P_T}{\bar{\boldsymbol{\Omega}}}^{-1}\right)^{-1}$ and ${\bf d} = -{\bf B}\bar{\bf s}$, problem \eqref{CI-WMMSE4} can be written as 
		\begin{align}
		\min\limits_{{\boldsymbol{\delta}}{\succeq}{\bf 0}}
		\left \|{\bf B}{\boldsymbol{\Lambda}}{\boldsymbol{\delta}}-{\bf d}\right\|_2^2.
		\label{CI-WMMSE NNLS}
		\end{align}
		It can be observed that ${\boldsymbol{\Lambda}}$ is a diagonal matrix with some zero diagonal elements when ${\bf s}$ includes QAM symbols. The zeros in the diagonal elements force the corresponding columns in ${\bf B}{\boldsymbol{\Lambda}}$ and elements in ${\boldsymbol{\delta}}$ to be zero during optimization. We denote ${\mathcal{T}}$ and  $K_{{\mathcal{T}}}$ as the index set of the non-zero diagonal elements of ${\boldsymbol{\Lambda}}$ and its cardinality, respectively. For different modulations, we have
		\begin{align}
		\mathbb{E}_{{\bf s}}\left\{K_{{\mathcal{T}}}\right\} = 
		\begin{cases}
		K\ \ {\rm if}\  {\bf s}\ {\rm is}\ {\rm modulated\ by}\ {\rm 16QAM}\\
		\frac{K}{2}\ \ {\rm if}\  {\bf s}\ {\rm is}\ {\rm modulated\ by}\ {\rm 64QAM}\\
		2K\ {\rm if}\  {\bf s}\ {\rm is}\ {\rm modulated\ by}\ {\rm PSK}\\
		\end{cases}.
		\end{align}
		We further define
		 ${\bf B}_{[:,j\in {\mathcal{T}}]}$/${\boldsymbol{\Lambda}}_{[jj\in {\mathcal{T}}]}$/${\boldsymbol{\delta}}_{[j\in {\mathcal{T}}]}$  as the matrix/diagonal matrix/vector composed of the columns/diagonal elements/elements of ${\bf B}$/${\boldsymbol{\Lambda}}$/${\boldsymbol{\delta}}$ whose indices belong to set ${\mathcal{T}}$. Then, problem \eqref{CI-WMMSE NNLS} is equivalent to
		\begin{align}
		\min\limits_{{\boldsymbol{\delta}}_{[j\in {\mathcal{T}}]}{\succeq}{\bf 0}}
		\left \|{\bf B}_{[:,j\in {\mathcal{T}}]}{\boldsymbol{\Lambda}}_{[jj\in {\mathcal{T}}]}{\boldsymbol{\delta}}_{[j\in {\mathcal{T}}]}-{\bf d}\right\|_2^2.
		\label{CI-WMMSE NNLS2}
		\end{align}
		Compared with \eqref{CI-WMMSE NNLS}, the dimension of ${\boldsymbol{\delta}}$ is reduced from $2K$ to $K_{\tau}$. {Our low-complexity optimization \eqref{CI-WMMSE NNLS2} is also suitable for CI-ZF with NNLS-based solution since they have the similar matrix structure  \cite{article, 8815429}, and these NNLS problems can be solved by the well-known active set based algorithm \cite{lawson1995solving}.}
		
		{Table \ref{complexity} compares the complexity (including computing the closed forms \eqref{solution 1} and \eqref{CIZF}) between CI-WMMSE and the optimal NNLS-based solution of CI-ZF in \cite{8815429}, which requires minimum complexity among several optimal solution forms of CI-ZF \cite{7103338,Li2021,8815429}. We omit the derivation of complexity because it only involves simple calculations. '(LC)' indicates that the NNLS problem is simplified by our low-complexity algorithmic solution.}  { $N_{L1}/N_{L2}/N_{L3}$ denote the numbers of main loops in active set based algorithm.  $\mathbb{E}\{M\}$ denotes the average $M$ with Rayleigh channel in SNR range from 0 to 40dB. $N_{M}$ is the number of multiplications that should be performed when a matrix with dimension $2N\times 2K$ multiplies a matrix with dimension $2K\times 2K$. As Table \ref{complexity} shows, our design in this subsection reduces part of the complexity, and CI-WMMSE has a similar low complexity to CI-ZF.}
		\begin{table}[hb]
			\renewcommand\arraystretch{1.2}
			\setlength\tabcolsep{3pt}
			\centering
			\caption{Complexity (Required Multiplications) Comparison}
			\begin{tabular}{cccc} 
				\hline
				\multirow{2}{*}{} & \multirow{2}{*}{
					{\begin{tabular}[c]{@{}c@{}}Required Multiplication\\ Number $M$\end{tabular}}} & \multicolumn{2}{c}{$\mathbb{E}\{M\}\  (N=K=12)$}                       \\ 
				\cline{3-4}
				&                          & \multicolumn{1}{l}{16QAM} & 64QAM                  \\ 
				\hline
				{{\begin{tabular}[c]{@{}c@{}}\textbf{CI-WMMSE}\\(LC)\end{tabular}}}         & {\begin{tabular}[c]{@{}c@{}}$\left(8N+4K_{{\mathcal{T}}} + 4 K_{{\mathcal{T}}}N_{L1}\right)K$\\$+(12+16N+13\frac{1}{3}K)K^2$\end{tabular}}                         & $4.09N_M$                         & $3.94N_M$   \\
				\hline
				\textbf{CI-ZF} \cite{8815429}             & {\begin{tabular}[c]{@{}c@{}}$(20K +
						8KN_{L2})N$\\$+(4+24N+4K)K^2$   \end{tabular}}                        & $4.17N_M$                          & $3.92N_M$  \\ 
				\hline
				{{\begin{tabular}[c]{@{}c@{}}\textbf{CI-ZF}\\(LC)\end{tabular}}}             &
				
				{\begin{tabular}[c]{@{}c@{}}$\left(4K_{{\mathcal{T}}} +12K +4 K_{{\mathcal{T}}}N_{L3}\right)N$\\$
						+(4+24N+4K)K^2$   \end{tabular}} 
				& $3.99N_M$                   &  $3.76N_M$                       \\
				\hline
			\end{tabular}
			\label{complexity}
		\end{table}
		\section{Numerical Results}
		In this section, we use the Monte Carlo method to evaluate the performance in the scenario of the Rayleigh fading channel. {Since the designs of ${\bf A}$ and ${\boldsymbol{\Omega}}$ are beyond the scope of the present work, we consider the special case of ${\bf A}={\bf I}$ and $\boldsymbol{\Omega} = {\bf I}$ during the simulation to highlight our contribution, based on which CI-WMMSE degenerates into CI-MMSE.} {It is assumed that
			a transmission block consists of $L$ groups of symbols, within whose time-frequency resource the wireless channel stays fixed.} {To facilitate the practical demodulation for the SLP schemes involved in the simulation, we unify the $\gamma$ in each transmission block by employing the power allocation scheme in \cite{9593254}.}
		
		\begin{figure}[htp]
			\centering
			\includegraphics[width=3.5in]{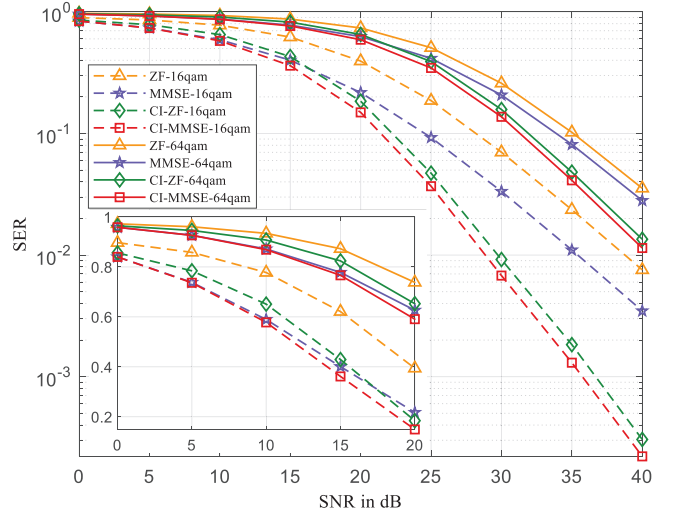}
			\caption{SER vs SNR, $N=K=12$, {symbol transmission ($L=1$)}.}
			\label{qam_ser}
		\end{figure}
		{Fig. \ref{qam_ser} shows the comparison of SER performance for $N=K=12$, in which the performance comparison in the low SNR ranges is magnified and displayed in the lower left. In order to be consistent with the most conventional SLP studies \cite{7042789,8466792,Li2021}, we consider $L=1$ in Fig. \ref{qam_ser}.} It can be observed that CI-MMSE achieves a significant gain than MMSE, and it also provides better performance than CI-ZF in full SNR ranges.
		\begin{figure}[t]
			\centering
			\includegraphics[width=3.5in]{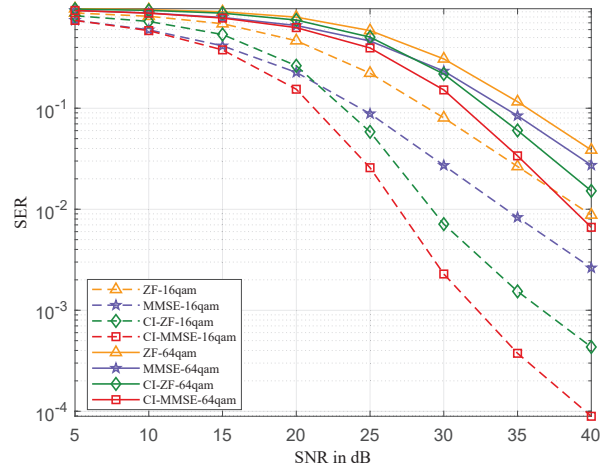}
			\caption{SER vs SNR, $N=K=12$, {block transmission ($L=1000$)}.}
			\label{qam_blkSer}
		\end{figure}
		The SER comparison with $L=1000$ is shown in Fig.  \ref{qam_blkSer}. CI-MMSE provides SNR gains of about 2.3dB and 7.4dB than CI-ZF and MMSE when the SER is $10^{-2}$ with 16QAM, and a similar trend can also be observed in 64QAM transmission. {There exist some performance differences between Fig. \ref{qam_ser} and Fig. \ref{qam_blkSer} since small $L$ has an impact on the SLP performance with power allocation \cite{9593254}.}

		Fig. \ref{12mul12_tp} compares the spectrum efficiency of the different techniques with the low-density parity check (LDPC) coding scheme \cite{1057683}. 
		{Since the received signals of the external constellation are extended and do not follow Gaussian distribution, the variance provided for the Gaussian soft demodulator is computed from the received signals of the inner constellation. We simulate the adaptive coding, which has not been studied in SLP, by decreasing the code rate from high to low to find the best-matched one for data transmission.} 
		Compared with CI-ZF, CI-MMSE provides about 37.8\% gain in spectrum efficiency when SNR is 15dB with 16QAM and 38.9\% gain when SNR is 20dB with 64QAM. It is worth noting that CI-MMSE provides better performance in full SNR ranges in both coded and uncoded systems.
		\begin{figure}[H]
			\centering
			\includegraphics[width=3.5in]{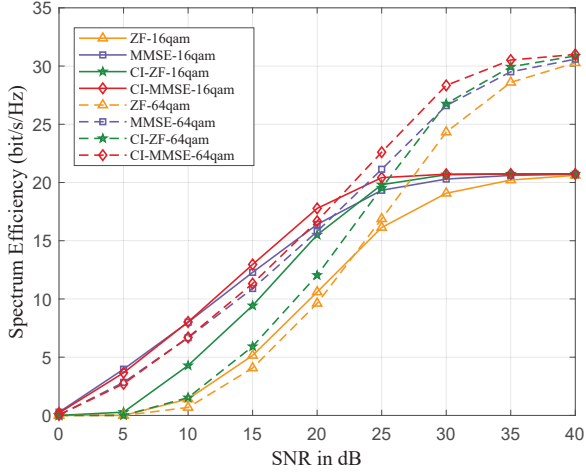}
			\caption{Spectrum Efficiency vs SNR, $N=K=12$, LDPC, $L=6000$.}
			\label{12mul12_tp}
		\end{figure}
		\section{Conclusion}
		In this paper, we propose the CI-WMMSE symbol-level precoding, which  minimizes the weighted mean square error between the received signal and the constellation point in CIR. In addition, we first put forward a simple generic description of CIR that facilitates subsequent SLP design. Based on the latter, the CI-WMMSE is further formulated as an NNLS problem that can be solved efficiently by the active-set algorithm. Besides, the WMMSE precoding  and the existing SLP solutions can be easily verified as special cases of the proposed scheme. The simulation results show that CI-WMMSE outperforms the state-of-the-art SLP schemes in full SNR ranges in both coded and uncoded systems {without additional complexity over conventional SLP}.

		% \ifCLASSOPTIONcaptionsoff
		% \newpage
		% \fi

		
		
		%\bibliographystyle{IEEEtran}
		% 参考文献出现横杠 D:\AAAA\TexLive\2019\texmf-dist\bibtex\bst\IEEEtran
		
		\bibliographystyle{IEEEtran}
		\bibliography{Refabrv,IEEEfull}

\begin{thebibliography}{10}
\providecommand{\url}[1]{#1}
\csname url@samestyle\endcsname
\providecommand{\newblock}{\relax}
\providecommand{\bibinfo}[2]{#2}
\providecommand{\BIBentrySTDinterwordspacing}{\spaceskip=0pt\relax}
\providecommand{\BIBentryALTinterwordstretchfactor}{4}
\providecommand{\BIBentryALTinterwordspacing}{\spaceskip=\fontdimen2\font plus
\BIBentryALTinterwordstretchfactor\fontdimen3\font minus
  \fontdimen4\font\relax}
\providecommand{\BIBforeignlanguage}[2]{{%
\expandafter\ifx\csname l@#1\endcsname\relax
\typeout{** WARNING: IEEEtran.bst: No hyphenation pattern has been}%
\typeout{** loaded for the language `#1'. Using the pattern for}%
\typeout{** the default language instead.}%
\else
\language=\csname l@#1\endcsname
\fi
#2}}
\providecommand{\BIBdecl}{\relax}
\BIBdecl

\bibitem{1040325}
A.~Bourdoux and N.~Khaled, ``Joint {TX-RX} optimisation for {MIMO-SDMA} based
  on a null-space constraint,'' in \emph{Proc. IEEE 56th Veh. Technol.Conf.
  (VTC)}, vol.~1, Sept. 2002, pp. 171--174.

\bibitem{974266}
H.~Sampath, P.~Stoica, and A.~Paulraj, ``Generalized linear precoder and
  decoder design for mimo channels using the weighted {MMSE} criterion,''
  \emph{{IEEE} Trans. Commun.}, vol.~49, no.~12, pp. 2198--2206, Dec. 2001.

\bibitem{1056659}
M.~Costa, ``Writing on dirty paper (corresp.),'' \emph{{IEEE} Trans. Inf.
  Theory}, vol.~29, no.~3, pp. 439--441, May 1983.

\bibitem{1413598}
B.~Hochwald, C.~Peel, and A.~Swindlehurst, ``A vector-perturbation technique
  for near-capacity multiantenna multiuser communication-part {II}:
  Perturbation,'' \emph{{IEEE} Trans. Commun.}, vol.~53, no.~3, pp. 537--544,
  Jan. 2005.

\bibitem{9035662}
A.~Li, D.~Spano, J.~Krivochiza, S.~Domouchtsidis, C.~G. Tsinos, C.~Masouros,
  S.~Chatzinotas, Y.~Li, B.~Vucetic, and B.~Ottersten, ``A tutorial on
  interference exploitation via symbol-level precoding: Overview,
  state-of-the-art and future directions,'' \emph{{IEEE} Commun. Surveys
  Tuts.}, vol.~22, no.~2, pp. 796--839, Mar. 2020.

\bibitem{Li2021}
A.~Li, C.~Masouros, B.~Vucetic, Y.~Li, and A.~L. Swindlehurst, ``Interference
  exploitation precoding for multi-level modulations: Closed-form solutions,''
  \emph{{IEEE} Trans. Commun.}, vol.~69, no.~1, pp. 291--308, Jan. 2021.

\bibitem{7103338}
C.~Masouros and G.~Zheng, ``Exploiting known interference as green signal power
  for downlink beamforming optimization,'' \emph{{IEEE} Trans. Signal
  Process.}, vol.~63, no.~14, pp. 3628--3640, Jul. 2015.

\bibitem{9593254}
A.~Li, F.~Liu, X.~Liao, Y.~Shen, and C.~Masouros, ``Symbol-level precoding made
  practical for multi-level modulations via block-level rescaling,'' in
  \emph{IEEE Workshop Signal Process. Adv. Wireless Commun. (SPAWC)}, Lucca,
  Italy, Sept. 2021, pp. 71--75.

\bibitem{7417066}
M.~Alodeh, S.~Chatzinotas, and B.~Ottersten, ``Constructive interference
  through symbol level precoding for multi-level modulation,'' in \emph{IEEE
  Glob. Commun. Conf., (GLOBECOM)}, San Diego, CA, USA, Dec. 2015, pp. 1--6.

\bibitem{8466792}
A.~Li and C.~Masouros, ``Interference exploitation precoding made practical:
  Optimal closed-form solutions for {PSK} modulations,'' \emph{{IEEE} Trans.
  Wireless Commun.}, vol.~17, no.~11, pp. 7661--7676, Sept. 2018.

\bibitem{8299553}
A.~Haqiqatnejad, F.~Kayhan, and B.~Ottersten, ``Constructive interference for
  generic constellations,'' \emph{{IEEE} Signal Process. Lett.}, vol.~25,
  no.~4, pp. 586--590, Apr. 2018.

\bibitem{8477154}
A.~Haqiqatnejad, F.~Kayhan, and B.~Ottersten, ``Symbol-level precoding design
  based on distance preserving constructive interference regions,''
  \emph{{IEEE} Trans. Signal Process.}, vol.~66, no.~22, pp. 5817--5832, Nov.
  2018.

\bibitem{4801492}
C.~Masouros and E.~Alsusa, ``Dynamic linear precoding for the exploitation of
  known interference in {MIMO} broadcast systems,'' \emph{IEEE Trans. Wireless
  Commun.}, vol.~8, no.~3, pp. 1396--1404, Mar. 2009.

\bibitem{7042789}
M.~Alodeh, S.~Chatzinotas, and B.~Ottersten, ``Constructive multiuser
  interference in symbol level precoding for the {MISO} downlink channel,''
  \emph{{IEEE} Trans. Signal Process.}, vol.~63, no.~9, pp. 2239--2252, May
  2015.

\bibitem{8462190}
Y.~Liu and W.-K. Ma, ``Symbol-level precoding is symbol-perturbed zf when
  energy efficiency is sought,'' in \emph{IEEE Int. Conf. Acoust., Speech
  Signal Process. (ICASSP)}, Calgary, AB, Canada, Apr. 2018, pp. 3869--3873.

\bibitem{article}
J.~Krivochiza, A.~Kalantari, S.~Chatzinotas, and B.~Ottersten, ``Low complexity
  symbol-level design for linear precoding systems,'' \emph{Proc.Symp. Inf.,
  Theory Signal Process Benelux}, pp. 1--8, 2017.

\bibitem{4712693}
S.~S. Christensen, R.~Agarwal, E.~De~Carvalho, and J.~M. Cioffi, ``Weighted
  sum-rate maximization using weighted {MMSE} for {MIMO-BC} beamforming
  design,'' \emph{IEEE Trans. Wireless Commun.}, vol.~7, no.~12, pp.
  4792--4799, Dec. 2008.

\bibitem{1391204}
C.~Peel, B.~Hochwald, and A.~Swindlehurst, ``A vector-perturbation technique
  for near-capacity multiantenna multiuser communication-part {I}: {C}hannel
  inversion and regularization,'' \emph{{IEEE} Trans. Commun.}, vol.~53, no.~1,
  pp. 195--202, Jan. 2005.

\bibitem{lawson1995solving}
C.~L. Lawson and R.~J. Hanson, \emph{Solving least squares problems}.\hskip 1em
  plus 0.5em minus 0.4em\relax SIAM, 1995.

\bibitem{8815429}
A.~Haqiqatnejad, F.~Kayhan, and B.~Ottersten, ``An approximate solution for
  symbol-level multiuser precoding using support recovery,'' in \emph{IEEE
  Workshop Signal Process. Adv. Wireless Commun. (SPAWC)}, Cannes, France, Jul.
  2019, pp. 1--5.

\bibitem{1049276}
M.~Joham, K.~Kusume, M.~Gzara, W.~Utschick, and J.~Nossek, ``Transmit wiener
  filter for the downlink of {TDDDS-CDMA} systems,'' in \emph{Proc. IEEE 7th
  Symp. Spread-Spectrum Technol.}, vol.~1, Applicat., Prague, Czech Republic,
  Sep. 2002, pp. 9--13.

\bibitem{1057683}
R.~Gallager, ``Low-density parity-check codes,'' \emph{IRE Trans. Inf. Theory},
  vol.~8, no.~1, pp. 21--28, Jan. 1962.

\end{thebibliography}
		
	\end{document}